\documentclass[11pt]{article}

\usepackage{a4}
\usepackage{color}
\usepackage{hyperref}
\usepackage{graphicx}

\begin{document}

\thispagestyle{empty}

\small
\begin{tabular}{lcr}
\large{\bf GEOS} & \hspace{0.8cm} \large{\bf GEOS CIRCULAR ON CEPHEID VARIABLES} & \large{\bf GEOS} \\
\large{\bf CEP 2} & {\bf }  & {\bf Parc de Levesville, 23} \\
{\bf } & {\bf July 2014}  &  {\bf F-28300 BAILLEAU L'EV\^EQUE} \\
\end{tabular}
\normalsize
\ \vspace{1cm}

\centerline{\large{ \underline{\bf GEOS RR Lyr Survey: FM DEL IS INDEED A CEPHEID } }}
\vspace{0.5cm}

\normalsize
{\bf Jean-Fran\c cois Le Borgne$^{1,2,3}$, Alain Klotz$^{1,2,3}$.} \\
$^1${GEOS (Groupe Europ\'een d'Observations Stellaires), 23 Parc de Levesville, 
28300 Bailleau l'Ev\^eque, France} \\
$^2${Universit\'e de Toulouse; UPS-OMP; IRAP; Toulouse, France} \\
$^3${CNRS; IRAP; 14, avenue Edouard Belin, F-31400 Toulouse, France}
\vspace{0.5cm}

\large{\underline{ABSTRACT}}
\vspace{0.5cm}

Though FM Del has been considered as a RR Lyr star by Preston et al. in 1959
(following discovery by Huth, 1957), Huth (1960) eventually changed his mind by
showing that it is in fact a cepheid of W Vir type of period of 3.95452 days.
Various authors since then have considered it as a cepheid indeed, with the exception
of Wils et al. (2006) who list this star in their RR Lyr catalog with a period of
0.79688 days. On this basis, FM Del was added to Tarot RR Lyr program. We
present here these observations which confirm the cepheid type.
\vspace{0.5cm}

\large{\underline{RESUM\'E}}
\vspace{0.5cm}

Quoique l'\'etoile FM Del ait \'et\'e consid\'er\'ee comme une RR Lyr par Preston et al. en 1959
suivant la d\'ecouverte par Huth (1957), Huth (1960) changea d'id\'ee en montrant
qu'il s'agit en fait d'une c\'eph\'eide de type W Vir de p\'eriode 3.95452 jours.
Plusieurs auteurs ont consid\'er\'e cette \'etoile comme une c\'eph\'eide depuis, \`a
l'exception de Wils et al. (2006) qui la listent dans leur catalogue de RR Lyr avec
une p\'eriode de 0.79688 jour. Sur cette base, FM Del a \'et\'e ajout\'ee au programme RR Lyr
de Tarot. Nous pr\'esentons ici ces observations qui confirment le type c\'eph\'eide.

\vspace{0.5cm}

\large{\underline{RIASSUNTO}}
\vspace{0.5cm}

Scoperta da Huth (1959), FM Del \`e stata classificata come una variabile di tipo 
RR Lyr da Preston et al. (1959). Successivamente, Huth (1960) ha mostrato che si 
tratta di una Cefeide tipo W Vir con periodo 3.95452 d. Questa classificazione \`e 
stata adottata da molti autori ad eccezione di Wils et al. (2006), i quali  la 
riportano nel loro catalogo di variabili RR Lyr con periodo 0.79688 d. Sulla base 
di questa nuova indicazione, FM Del \`e stata aggiunta al programma RR Lyr svolto 
con Tarot. L'analisi delle  nuove osservazioni conferma definitivamente che si tratta 
di una cefeide.

\vspace{0.5cm}

\large{\underline{RESUMEN}}
\vspace{0.5cm}

La estrella FM Del fue clasificada como de tipo RR Lyr por Preston et al. en 1959, 
tras su descubrimiento por Huth (1957). Posteriormente, Huth (1960) demostr\'o que se 
trata de una Cefeida de tipo W Vir, con periodo de 3.95452 d\'ias. Desde entonces numerosos 
autores la han considerado como Cefeida, con la excepci\'on de Wils et al. (2006), que la 
incluyeron en su cat\'alogo de variables de tipo RR Lyr, con un periodo de 0.79688 días. 
En base a esto, FM Del ha sido a\~nadida al programa de RR Lyr de Tarot. Presentamos 
aqu\'i estas observaciones, que confirman que es una Cefeida. 

\vspace{1cm}

\section{Introduction}
As recalled by Diethelm (1986), FM Del was discovered by Huth (1957) as an RR Lyr
of period 0.79739 days and was studied as such by Preston (1959). Preston's study
concerned metallicity which was found to be about solar. This pushed Huth to
reconsider the type determination and eventually revised (Huth, 1960) its type to
cepheid with period 3.95542 days.
GCVS (Samus et al., 2007-2012) gives CWB type (population II cepheid) and variation 
from 12.3 to 13.3 (p) with reference to Huth (1960).\\
Since then, several authors used FM Del as a cepheid. It was listed cepheid general
studies by Petit (1960a, 1960b) and Harris (1985). New observations only appear
in Diethelm (1986, 1990) who made a single photometric measurement in Walraven
VBLUW system determining a metallicity index [Fe/H]=-0.9. Having only one
measurement, Diethelm supposed that it is a cepheid, but with caution. Schmidt
et al. (2003) eventually published electronic measurements in V and R filters,
followed by spectroscopic observations (2005). Though Schmidt et al. assume the same
period as determined by Huth (1960), their light curve leaves no doubt on the
cepheid type.\\
Strangely enough, FM Del appears in Wils et al. (2006) catalog of RR Lyr stars
from Rotse measurements with period of 0.79688 days (NSVS 11462023), a period
close to Huth's. We may guess that this is an artifact from the time sampling
of one measurement per night. It is worth to note that the ratio between the two
periods is close to 5 (4.96). A consequence of FM Del being in Wils et al. catalog
have been to schedule FM Del in Tarot RR Lyr program.\\
\section{TAROT observations}
A description of TAROT telescopes may be found in Klotz et al. (2008) and GEOS
RR Lyr survey, including TAROT RR Lyr program in Le Borgne et al. (2007, 2012).
\begin{figure}[ht]
\centerline{\begin{tabular}{lcl}
   \includegraphics[width=9cm]{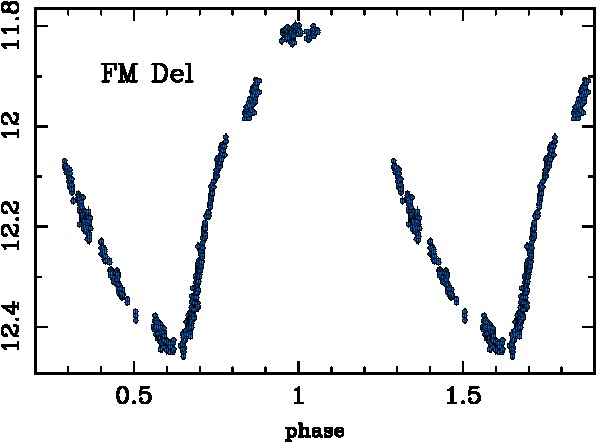}    &
   \hspace{0.5cm}   &
   \includegraphics[width=7cm]{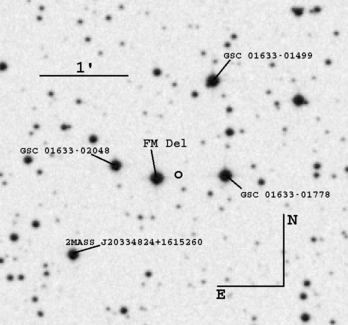}  \\
Figure 1: Folded light curve of FM Del (TAROT) \\using elements (\ref{elements}). &
   \hspace{1cm}   &
Figure 2: Star field around FM Del.\\
   \end{tabular}}
\end{figure}
To summarize, let us say that TAROT telescopes are robotic 25cm telescopes aimed
to the observation of optical counterparts of events triggered by $\gamma$-ray 
satellite alerts or astro-particles detectors (neutrinos, gravitational waves). 
One is located
in France (Calern Observatory) and the other in Chile (La Silla Observatory).
Between alerts, the telescopes are used for several programs, one of them being
to contribute to GEOS RR Lyr survey.\\
FM Del has been observed by the northern TAROT telescope at Calern Observatory from JD
2455401.414 (23  July 2010) to 2456162.582 (23 August 2012). 872 measurements spread
over 29 nights have been obtained. As for the other stars of TAROT RR Lyr program, 
data reduction, from bias subtraction and flatfielding to photometry using SExtractor 
(Bertin and Arnouts, 1996), is performed automatically. \\
In TAROT RR Lyr program, observations are scheduled in order to obtain times of
maximum during selected nights. For FM Del, the nights where selected according
the elements given in Wils et al. (2006). It appeared that no maximum was observed
during the selected nights, and furthermore, the star varied very few during all
of them, and at different mean brightness. This is not typical of a RR Lyr star.
We then plotted a folded light curve with the period given by Huth and later
used by Schmidt et al. (Figure 1). We first used the elements given in GCVS (Samus
et al., 2011) but the maximum of the light curve did not correspond to phase 0.
We then adjusted the origin of the elements: \\
\section{Discussion}
One fact is to be noted: there is no reference to Wils et al. (2006) in CDS/SIMBAD
entry for FM Del, nor NSVS 11462023 is given as cross identification. The question
is then to investigate if these are 2 different stars. The coordinates of FM Del
given in GCVS are the same as those given in SIMBAD. The coordinates of NSVS 11462023
from ROTSE are 0.2 arc minutes from GCVS FM Del coordinates.
Figure 2 shows the star field around FM Del. The circle shows the position of NSVS 
11462023 from Wils et al. (203343.42+161619.2). There is no star at 
ROTSE position and the closest 12th magnitude star is FM Del. Then it is most probable 
that FM Del and NSVS 11462023 are the same star.\\
\begin{equation}HJD\ 2456111.65 + 3.95452\ E.\label{elements}\end{equation}
\begin{figure}[ht]
\centerline{\begin{tabular}{lcr}
   \includegraphics[width=8cm]{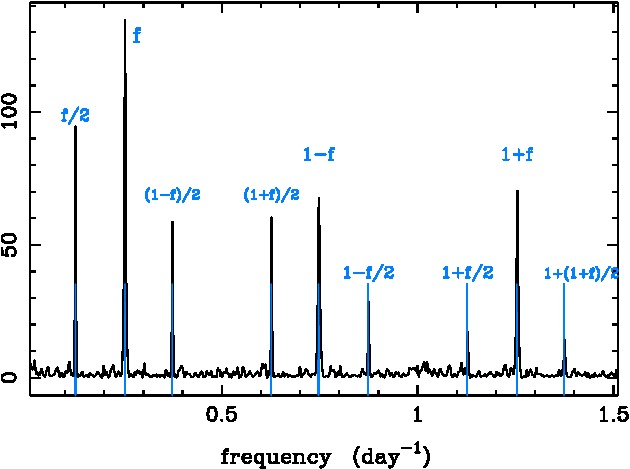}      &
   \hspace{0.3cm}                                        &
   \includegraphics[width=7.7cm]{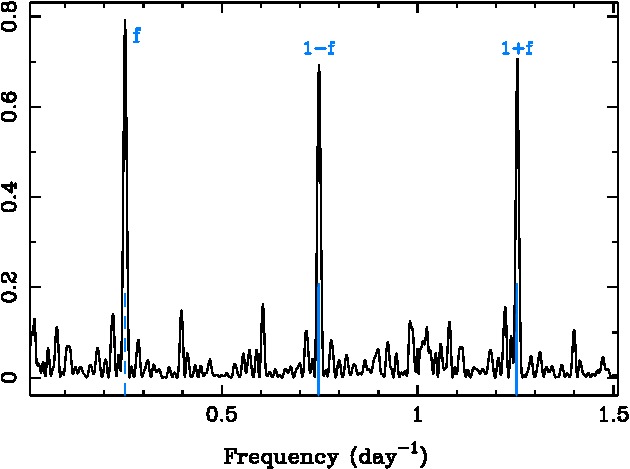}  \\
   \multicolumn{3}{c}{Figure 3: Periodograms of Rotse measurements.
   left: Schwarzenberg's method. Right: Vani\^cek's method.}   \\
\end{tabular}}
\end{figure}
Next step is to check by our self the frequencies present in Rotse measurements
which are available at the web site http://skydot.lanl.gov/ (Wozniak, 2004).
Rotse database
contains 103 measurements for NSVS 11462023 between JD 2451420.746 (30 August
1999) and 2451511.631 (29 November 1999). A periodogram of these data (Figure 3,
left), using multiharmonic Fourier series method (Schwarzenberg-Czerny, 1996),
gives a main frequency $f$ at 0.25307 d$^{-1}$, that is 3.95147 days. We also see
aliases at $1-f$ and $1+f$, as well as their half frequencies $f/2$, $(1-f)/2$ and 
$(1+f)/2$. Other aliases appear at $1-f/2$, $1+f/2$ and $1+(1+f)/2$.
Note that because $f$ is close to 0.25, $5f$ is close to $1+f$. As noted above, 
5 is the ratio between the GCVS cepheid period and the period given by Wils et al..
Obviously, they used the alias $1+f$ as the main frequency. 
In principle, since the time sampling of Rotse is about one measurement per day,
they should have considered such a frequency only with caution.

We also use a second method, the iterative sine--wave least--squares method
(Vani\^cek, 1971) to build a second periodogram (Figure 3, right)
the frequency $f$ found is 0.25274 d$^{-1}$, corresponding to a period of 3.95663
days. As with the former method, we see aliases at $1-f$ and $1+f$ which is expected 
with the time sampling of the measurements. However, none of the other aliases 
appears. \\
\begin{figure}[ht]
\centerline{\begin{tabular}{c}
   \includegraphics[width=8cm]{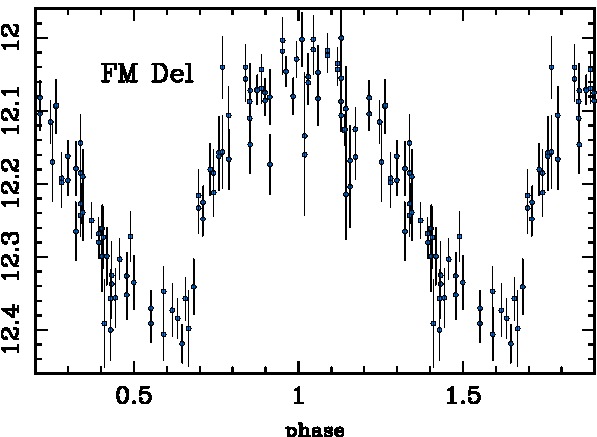}\\
   Figure 4: Folded light curve of Rotse measurements using elements (\ref{elements}).\\
\end{tabular}}
\end{figure}
The folded light curve of Rotse measurements are plotted with the
same elements as TAROT measurements in figure 4. \\
As a comparison, periodograms of Tarot data obtained with the two methods are given
in figures 5. Thanks to TAROT time sampling and duration, only $f$ and $f/2$ appear 
in periodogram obtained with Schwarzenberg-Czerny's method while only $f$ is found 
with Vani\^cek's method. However, Vani\^cek's periodogram is noisier. The remaining 
lines are aliases of 1 d$^{-1}$ frequency. 

All these periodograms, on Rotse and Tarot data, give an uncertainty on period of 
about 0.003 day. This does not allow to improve the period given in GCVS.
\begin{figure}[ht]
\centerline{\begin{tabular}{lcr}
   \includegraphics[width=8cm]{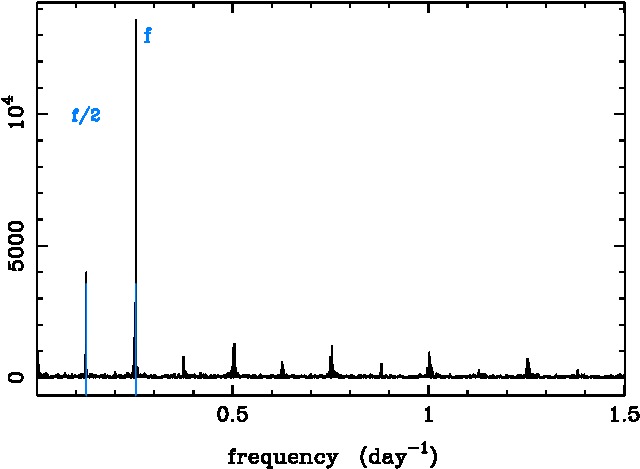}      &
   \hspace{0.3cm}                                        &
   \includegraphics[width=7.7cm]{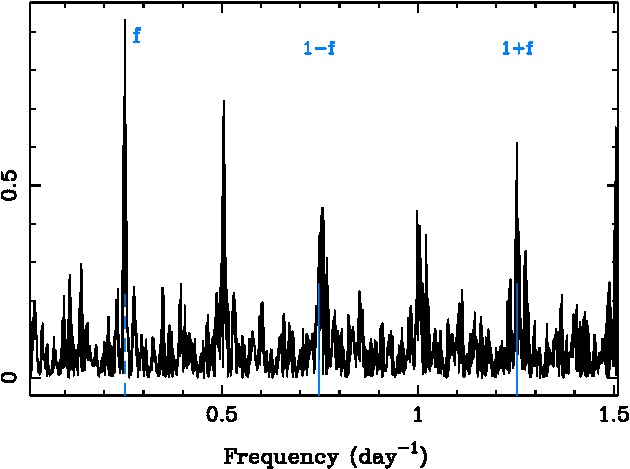}  \\
   \multicolumn{3}{c}{Figure 5: Periodograms of Tarot measurements.
   left: Schwarzenberg's method. Right: Vani\^cek's method.}   \\
\end{tabular}}
\end{figure}
Note that FM Del is not in Catalina Survey database.
\vspace{5cm}

\section{Conclusion}
We have confirmed that FM Del is a cepheid as it was supposed to be since 1960
 although it was first identified as an RR Lyr at its discovery in 1957.
It was erroneously added to Wils et al. (2006) RR Lyr catalog with a period which
is 5 times less than true one. Included in RR Lyr TAROT program on this basis,
it clearly appeared not to be a RR Lyr star. TAROT data fit nicely Huth's 1960
period, as do ROTSE measurements which were at origin of Wils et al. paper.

\end{document}